# Negative Human Rights as a Basis for Long-term AI Safety and Regulation[1]

**Ondrej Bajgar**   ONDREJ@BAJGAR.ORG
*Department of Engineering Science & Future of Humanity Institute*
*University of Oxford, United Kingdom*

**Jan Horenovsky**   HORENOVJ@PRF.CUNI.CZ
*Department of Politology and Sociology, Faculty of Law*
*Charles University, Czech Republic*

## Abstract

If autonomous AI systems are to be reliably safe in novel situations, they will need to incorporate general principles guiding them to recognize and avoid harmful behaviours. Such principles may need to be supported by a binding system of regulation, which would need the underlying principles to be widely accepted. They should also be specific enough for technical implementation. Drawing inspiration from law, this article explains how negative human rights could fulfil the role of such principles and serve as a foundation both for an international regulatory system and for building technical safety constraints for future AI systems.

## 1. Introduction

As human-made systems become more and more autonomous — able to act without human guidance in an ever wider array of situations — we will need to equip them with *general* principles to decide which actions or outcomes are desirable to humans and which should be avoided. A narrower set of instructions risks guiding a system toward solutions that fulfil such instructions but oppose wider human values. For instance, we have already seen claims that recommender systems may be pushing users toward more extreme views, which helps the algorithms fulfil their narrow goal of maximizing time spent on a website by making the users more predictable but violates wider human preferences over what the algorithm should be doing (Russell, 2019).[2] As AI systems continue becoming more capable, the costs of such specification failures are likely to grow. Thus, there have been calls for advanced AI systems to be taught human values or preferences (Yudkowsky, 2011) in order to solve what has been called the *AI alignment problem.*

However, what values should these be? The phrase *human values* obscures the wide disagreements among humans over their values and preferences. That becomes a serious issue if we consider a need for binding regulation, which is likely to be necessary for globally addressing negative externalities from AI, including the one mentioned for recommender systems but possibly going all the way to existential risks to humanity (Bostrom, 2014; Petit, 2017, 2020; Critch & Krueger, 2020; Ngo, 2020). If regulation is to require a set of principles or values to be respected, it needs to (at least indirectly) specify which values, and if such a regulatory framework is to be accepted and enforced internationally, we need a set of values that are themselves widely endorsed by the international community.

---

[1] This is a copy of the final version of the article as published in the *Journal of Artificial Intelligence Research* 76 (2023) 1043-1075, which can be found at https://jair.org/index.php/jair/article/view/14020 or https://doi.org/10.1613/jair.1.14020.

[2] See Krakovna et al. (2020) for a collection of many other examples of artificial intelligence (AI) specification gaming.

This article proposes a solution to this double requirement of values that are at the same time general and widely endorsed. It suggests using the concept of *negative human*[3] *rights* as a minimal set of principles that all AI systems should respect and that are already widely accepted and protected. We restrict the proposal to negative[4] rights since requiring all AI systems to help fulfil positive rights does not seem sensible (the state being the natural duty-bearer there). Among negative human rights, we propose initially concentrating on those for which there is already wide consensus, such as the right to life or the protection of property.

To give concrete meaning to this seemingly abstract concept, we draw inspiration from law, where the judicial system is designed to give a single authoritative answer to whether any particular behaviour is permitted — this already largely protects human rights, through both international and national law and judicial systems. In a similar vein, the proposal sees a behaviour to be in accordance with human rights if and only if a particular system of adjudication would judge it to be so[5] based on a convention describing what specific rights are protected.

In practice, the adjudication system would be directly invoked only in a small minority of cases — the aim is to design future AI systems to themselves predict which behaviours may violate negative human rights and eliminate such plans as options before they are executed. Our proposal naturally extends to an associated technical task of classifying any particular (hypothetical) behaviour as compliant or not, with the adjudication system as a source of ground truth — a kind of task suited for the machine learning (ML) paradigm. Thus, the proposal naturally suggests both governance structures to keep AI safe in the long term and a framing of the associated technical problem.

Our article makes the following contributions:

- We provide a new framing for the AI alignment problem, emphasizing the need for separating positive goals and safety constraints. In line with that, we focus on how an AI system should not behave, not on the whole question of how it should behave.
- Drawing on insights from legal theory, we propose negative human rights as the value foundation for those safety constraints and explain why they are well suited for this purpose.

---

[3] While we are using the term *human* rights, the rights of other subjects, such as non-human animals (Owe & Baum, 2021) or digital minds (Shulman & Bostrom, 2021), could be included within the framework as we go forward. This would be in line with the trend of an extending sphere of subjects of fundamental rights and, relatedly, the expansion of the moral circle (Anthis & Paez, 2021). The key principle we are building on is consensus, which currently does not globally include protections of non-human sentient beings, perhaps beyond basic protection against animal torture in some countries. Thus, we are using the term human rights (and sometimes humans) to match the current historical context and, in particular, are not trying to make a normative statement that rights of non-human rights should be excluded in the long run. Indeed, non-human subjects have already appeared in some AI-related declarations, such as the *Montreal Declaration for a Responsible Development of Artificial Intelligence* (2018). Similarly for rights of yet non-existent subjects, such as future generations.

[4] Negative rights, which our proposal centres on, protect their subject from harmful outside interference, as opposed to positive rights, which entitle their holder to something, e.g., education. More on this in Section 3. When we refer to human rights throughout the paper, we generally mean negative rights only, unless stated otherwise.

[5] Such a position is related to legal realism (Fisher, Horwitz & Reed, 1993) — rather than inquiring into the theoretical meaning and nature of law, we are trying to predict law as it is actually practised in reality.

- We explain how compliance with negative human rights could be used to build technical solutions for safe AI and outline some of the associated challenges as inspiration for future technical work, as well as some existing areas such work could build on.
- Relative to what we see as the default way of doing regulation — an array of narrower rules targeting specific problems that have occurred — our proposal offers more general guidance to AI systems with a better chance of scaling into unforeseen future situations.
- We provide an early exploratory proposal for how an international regulatory system based on human rights could work to illustrate the previous points more concretely.

We would like to point out the following three limitations:

- We provide only a partial contribution to solving the specification problem in designing safe future AI, which may need to be complemented by suitable positive goals (in the direction of *intent alignment*) as well as other safety precautions, including legal ones.
- We do not provide concrete technical solutions but rather a framing that sets human rights compliance as a machine learning task amenable to future technical work.
- In some sections, we assume strong capabilities of future advanced AI systems, especially in terms of predicting the impacts of their actions. While capabilities of AI systems are rapidly advancing, their concrete form and degree partly remain a matter of speculation, so some of our arguments remain correspondingly uncertain.

After reviewing some existing work, we will first briefly explain which AI systems we are targeting. Then, in Section 3, we outline our conception of human rights for this purpose and how it parallels their use in law. We proceed in Section 4 by outlining some of the technical considerations for negative-human-rights-based AI safety and, in Section 5, describe an early exploratory proposal for how this could be integrated into a governance system.

**1.1 Existing Work**

Our work relates mainly to three strands of existing work: (1) academic work discussing AI and human rights, (2) lists of general principles supposed to guide AI development, and (3) legal and policy documents addressing human rights issues related to AI systems. We will now briefly discuss the three in turn, while for now omitting technical literature, some of which is referenced in Section 4 after readers are exposed to necessary context.

In the first strand, human rights have already appeared in discussions on the regulation of AI in academic literature. Articles in that space can be classified, on the one hand, according to whether the article considers only the near-term perspective (mostly pointing out problems with currently deployed technologies) or whether it also considers the long term, including more advanced future systems. On the other hand, we can look at whether the work also considers technical solutions or remains on the abstract level or the level of interpreting real-world effects of AI through the human rights lens.

Most work has been focusing on already existing problems — work casting many of those problems as human rights issues includes Latonero (2018), Cowger (2020), Anderson (2018), Rodrigues (2020), Aizenberg & Hoven (2020), and Raso et al. (2018). We find this strand of work particularly useful in highlighting the relevance of human rights to AI and in casting a large part of the many problems with AI under a single framework. Much work has also been done on solving technical problems related to present-day violations of human rights — work on preventing bias and discrimination being the most prominent example (Mehrabi et al., 2021). We are, however, trying to move beyond reacting to problems already present toward a forward-looking approach. Also, on the technical side, we are trying to move from a patchwork of solutions to particular problems toward a more general umbrella solution. For a long time, the particular solutions are likely to remain superior in their respective domains, but there is a risk of new problems falling through the cracks between the specific approaches — for this reason, we think the ambitious and difficult task of finding a more general solution is worth the effort.

There is work that, in the context of AI, considers human rights in their full generality. For instance, Donahoe & Metzger (2019) defend them as successfully accomplishing what other ethical frameworks for AI are trying to achieve — put the human being at the centre, cover a wide range of concerns, and rest on broad global consensus. We are also building especially on the latter two, while the first is interesting in its focus on the actual consequences of AI systems' behaviour, rather than the particular mechanisms by which a harm is caused (e.g., biased data) and which may be extremely difficult to exhaustively predict in regulation. Other authors mentioning human rights in a long-term context include Risse (2019) and Gabriel (2020). But while they appreciate the guidance human rights provide on the governance side, they do not make the bridge towards technical solutions that we are attempting to make here. Similarly, we feel the technical side is almost completely neglected in Fjeld & Nagy (2020), Anderson (2018), and Rodrigues (2020). We are trying to provide a framework that covers both the governance and the technical sides of the problem in a way that could, in its essence if not its details, persist in the long term.

This fact that human rights appear in both the shorter-term and long-term contexts suggests they could help in bridging these two perspectives. If robust human rights protection helps with both near- and long-term threats, human rights could form a basis for an *incompletely theorized agreement*, as has been called for by Stix and Maas (2021), making the two perspectives converge not necessarily on a philosophical level, but on the level of practical solutions.

McGregor et al. (2019) propose using *international human rights law* in particular as a framework for *algorithmic accountability* in a paper very close in spirit to our proposal — though they take as their point of departure work on *algorithmic transparency*, while our emphasis is on long-term AI safety. We think the contents of our papers are compatible and provide almost disjoint, complementary content. Furthermore, while existing international human rights law focuses mainly on states as duty bearers, we emphasize mainly duties of the AI-developing organizations and mechanisms within AI systems themselves — a needed shift emphasized also by Smuha (2021), for example.

The closest perspective to ours, though with stronger philosophical emphasis, is taken by Gabriel (2020), who gives three propositions related to long-term AI alignment relevant to our proposal. Firstly, similarly to us, he argues that the normative and technical aspects are interrelated. Secondly, he states that it is vital to be clear about the goal of alignment and distinguishes between minimalist and maximalist conceptions. Our proposal fits into the minimalist approach in this division as it only avoids unsafe outcomes, rather than trying to maximize positive value from AI. Thirdly, he states that it is necessary to identify the correct principles or values that alignment can rely on. He identifies human rights as one option, which we elaborate on here in much more detail.

In middle ground between the maximalist and minimalist conception, the field of *Cooperative AI* (Dafoe et al., 2020, 2021) seeks to build AI that helps achieve mutually beneficial outcomes in a world wherein goals of actors differ. We try to ensure the minimalist part of this objective — that the actors do not cause harm to each other — as something more clearly amenable to direct regulation. However, the broader goals of Cooperative AI are a useful extension that we certainly consider worthy of pursuing.

In the second strand of literature, beside the above academic work, many international organizations, governments, and private companies have published lists of general principles of AI over the past few years, many of which relate to human rights — Jobin et al. (2019) provide a good review of published lists. Many, for example the recent UNESCO Recommendation on the Ethics of AI (2021), mention human rights directly. Others mention principles that could easily be interpreted as falling within the scope of negative human rights — Jobin et al.'s paper found that 60 out of 84 lists of principles contained non-maleficence (under which they also grouped, for example, security, safety, or protection from harm) and 47 included privacy — both of which can be interpreted as subsets of human rights. Furthermore, 60 lists contained responsibility (a category including accountability and liability, which we are also trying to emphasize with respect to human rights specifically). From our perspective, many of those lists are shorter-term focused compared to the present article, and almost all lack detail on what the terms contained therein mean, not to mention a lack of detail on how they should be put into practice both on the technical and on the governance sides — things we are trying to address in much more detail than is usual in the documents on principles. We are in fact proposing to replace many of those principles with human rights, as they provide a more consistent, more complete, and more general framework ready for challenges that may arise in the future.

Finally, the issue of potential threats to human rights from AI systems is also making its way into policy papers and regulatory proposals. Under European law, the *Artificial Intelligence Act* (EU, 2021) considers the protection of human rights as one of the main reasons why new regulation is needed. In the US environment, the threat to human rights as a problem is mentioned, for example, in the *Final Report of the National Security Commission on Artificial Intelligence* (2021). The Australian Human Rights Commission is moving in a similar direction in its *Human Rights and Technology Final Report* (2021), which addresses the full range of potential issues related to AI and human rights. However, these works often omit how human rights protection could scale in the long term and do not consider how the requirement to respect human rights could be translated into concrete technical solutions — gaps we are trying to fill.

## 2. Artificial Intelligence

This article primarily concerns the long-term safety of artificial intelligence — any of a broad family of man-made systems autonomously performing tasks that people would generally associate with intelligence. AI is a diverse group of technologies, so let us outline, on the one hand, which technologies our framework is able to accommodate, and, on the other, what kinds of systems it is specifically targeting and would be most useful for.

On the most general level, it seems sensible to require that no technologies, for which that may be a concern, infringe on negative human rights — for instance, do not harm humans, damage their property, or limit their freedom.[6] The general objective of this proposal — AI respecting negative human rights — is thus relatively agnostic to where we draw the line of what already counts as an AI system, and it could include present-day systems such as autonomous vehicles, robotic cleaners, or recommender systems. In fact, there have already been calls for using the framework of human rights to assess the risks of present-day systems (Latonero, 2018). The advantage of this general interpretation is that the governance elements of the framework could start being developed relatively soon, even though on the technical level, traditional engineering methods for ensuring safety could continue being used in the near term, and incremental changes to current governance structures (e.g., tort law) could also suffice (in fact, many infringements on human rights via technologies would already be illegal under current legal provisions). Still, there may be advantages to implementing the more general structures suggested here *before* the more advanced technologies are put in place (Cave & ÓhÉigeartaigh, 2019).

Our primary concern is systems further in the future, which may operate in a broader set of situations (thus requiring more general decision principles) and with much larger solution spaces (thus being able to come up with solutions harder for humans to predict). Consequently, we will need more general measures to avoid undesirable outcomes both on the technical level (the decision-making of the AI systems themselves) and the governance level (to provide incentives also for governments and AI-developing organizations to implement effective safety precautions[7] in a way that scales with the growing generality of AI systems, beyond what the current, often narrow, legal rules can cover).

Though general principles can provide helpful guidance, not all AI systems would have the capacity to learn to recognize negative human rights in full generality, and it would not even be needed for all of them. The level of rights understanding required for a specific system would need to be appropriate for its degree of autonomy and the particular risks it would pose. However, we hypothesise that if we build systems above a certain level of generality and autonomy in the future, full understanding and compliance with negative human rights would be desirable. This proposal is written with such a situation in mind — so when we talk about *AI* or *AI systems*, this is what the reader should primarily imagine — but many elements of the proposal are also applicable to lesser systems through appropriate partial solutions.

This proposal focuses on *civilian* uses of AI systems, where the goal can be to eliminate infringements on negative human rights entirely. If allowed, military and other security deployments of AI systems are a more complicated case, which we are setting aside for now and is discussed elsewhere (Petit, 2018; Warren & Hillas, 2017). However, many parts of the solution frameworks proposed here could be suitably adapted to be useful even in non-civilian AI systems — for example, by replacing negative human rights with elements from (suitably extended) international humanitarian law — which we leave for future work.

---

[6] This is in line with the fact that general principles of legal protection are concerned with preventing societal harms of any kind — on such a general level, AI or other emerging technologies do not have any special status (see also Bennett Moses, 2017).

[7] This may also include questions of legal liability of AI-developing organizations.

## 3. Human Rights

We will now explain the conception of human rights this proposal adopts and how it is inspired by their use in law.

### 3.1 What Are Legalistic Human Rights

From a legal perspective, a human right is a kind of normative sentence — a sentence stating that something "ought to be" (Kelsen & Hartney, 1991). For instance, Article 5 of the European Convention on Human Rights states that "*everyone has the right to liberty and security of person*", meaning that no one's liberty or security ought to be infringed without legal permission, and defining a general basis for deciding concrete cases.

However, on its own, this is too vague and abstract, as described by Bellamy (2007, pp. 50-51): "*Charters are necessarily phrased at a high level of abstraction in order to promote agreement. That leaves the disagreements over the substance, scope, sphere and subjects of rights, the best means to secure them, and so on still to be decided. [...] Even those rights which reflect what practically everyone recognizes as great moral wrongs, such as murder or rape, can still involve disagreements over how they should be interpreted and interact with other rights.*"

Thus, to understand the material content of human rights, we need to look beyond abstract words at the effect of rights on the real world: which behaviours actors decide to avoid with respect to the rights holder. On the legal side, that should roughly correspond to what an adjudication system — typically courts of justice — would judge as an unlawful violation of a human right under particular circumstances. Instead of excessively dwelling on the definition of human rights as such, it may be enough to identify which behaviours are and which are not permissible. Since the functioning of adjudication systems is already a well-established and well-tested way of interpreting abstract concepts like human rights in practical situations, the meaning of human rights we use in the rest of this proposal defers to an adjudication system (which we elaborate on more in Section 5) as follows:

*A behaviour (actual or hypothetical) is in accordance with human rights if a particular adjudication system would judge it to be so.*

How this process works and how it is justified is studied in legal theory and is beyond the scope of this article, though three questions usually have to be answered in each case of assessing whether human rights have been infringed: (1) which concrete human rights are relevant to the given situation? (2) does the specific conduct of AI interfere with these rights? and (3) if so, is the interference (based on the internal rules of the normative system) justified on legally permissible grounds? The third mentioned point implies that not every interference with a human right is *per se* illegal (in other words, an infringement or violation) — for instance, the right to free speech sometimes gets limited in the interest of national security.

From this perspective, human rights law aims to be a rational system designed to yield a single authoritative final answer. The legal world offers a binary view of behaviour or its consequences — either legal or illegal, permitted or banned. Particular behaviour either unlawfully interferes with one's human rights or does not. In law, at least during the final decision in the courtroom, there is no possibility of a "maybe". Human rights law can thus be understood as a body of systemised prediction (Holmes, 1897). It is about predicting a potential decision in a specific situation based on statutes or conventions, past decision-making practice, and other relevant learning sources. From this perspective, actors like individuals or companies evaluate whether something is legal by predicting (hypothetical) court decisions.

This structure of the legal and judicial system could make it suited for machine learning (Benjamin, 2016; Casey & Niblett, 2017). Based on thousands of existing cases and literature from different jurisdictions and legal environments, AI may be able to learn to estimate the probability that its actions would violate any of the provisions in a particular convention (though other learning sources would also be necessary — more on this in Section 4).

### 3.2 Which Rights Should be Protected and Why

Legal theory distinguishes two kinds of legalistic human rights: positive and negative (Jellinek, 1892). The essence of positive rights is that something should be provided to someone — it can, for instance, create an obligation for the state to pay social benefits. Conversely, the essence of negative rights is the creation of a sphere of personal autonomy, which other actors must not violate (Melro & Oliveira, 2019, p. 140). In other words, negative rights create an imaginary protective barrier against interference.

The system that this article is proposing is based on *negative* human rights. Our proposal suggests that a subset of *negative* rights should bind every AI system globally to protect the most important aspects of our lives.[8] Firstly, while for positive rights, the state as a public authority is the natural duty-bearer, negative rights, in their broadest interpretation, may protect each individual from unjustified interference from any actor, including AI systems. We do not dispute that it makes sense to seek assistance from AI with providing for positive rights; however, this does not seem applicable to *all* AI systems and may better be encouraged through other mechanisms than the strong protections of negative rights that we want to propose here. There is no reason why *all* AI systems should be obliged to provide, for example, health care or pension. Our goal is to ensure that AI systems do not infringe on people's autonomous sphere, not to determine what AI systems should do as their positive function.

Besides, negative rights seem to be less controversial in the international community than many positive rights, like economic and social rights (Courtney, Hirschl & Rosevear, 2014). And while there may not be a complete consensus on negative human rights as protecting citizens from their states in today's world (Law & Versteeg, 2011), we believe that finding a consensus on negative human rights in relation to AI systems could be much more accessible. No one wants dangerous AI autonomously violating human rights in their streets, including authoritarian and undemocratic regimes.

Our proposal offers a shift in understanding human rights, which have been historically seen, within human rights law, mainly as protecting citizens against the state, while we are extending them to protect against actions taken by AI. In some, mainly European, jurisdictions, the horizontal effect of constitutional human rights (Gardbaum, 2003) already enables their direct application to non-state actors as duty bearers. In line with this practice, an extension to AI is natural and arises from the increasing need to protect humans from potentially powerful autonomous machines.

---

[8] Note that many prima facie positive rights include a negative aspect. So, while a right to health care may primarily imply an obligation for the state to provide health care (a positive right), it also includes a negative right that no one should interfere with one's access to health care. This negative aspect may need to be carried over to AI systems in some form, but again, we would suggest this only in cases where there is consensus, which may be less frequent for positive rights than for purely negative ones.

Which particular rights would be protected would need to be specified in a document such as an international convention and would have to result from a political deliberation process. However, some examples of which rights would almost certainly be protected — and already widely are — are the right to life, security of the person, the right to property, and the right not to be tortured.[9] For a more extensive list of rights, see, for example, the negative subset of the European Convention on Human Rights (ECHR, 1953) or the Universal Declaration of Human Rights (UN, 1948). Remarks on implementing the corresponding convention are provided in Section 5.

The protection of most of these human rights has much deeper roots than the concept of human rights as such. What we mean by protecting rights largely serves the same objectives that criminal law has served for centuries across the world, and what is thus less emphasized under human rights law for historical reasons. But where criminal law is usually phrased in a fairly specific, precise way, we think pointing to the general values it is protecting (which we think largely coincide with negative human rights) provides much better robustness in unforeseen future situations.

New rights might need to be added, once consensus is reached — for instance, Russell (2019, pp. 105-107) suggests a *right to mental secur*ity, which includes, for example, a right not to be lied to by AI systems. Relatedly, Ienca & Andorno (2017) suggest the rights to *cognitive liberty*, to *mental privacy*, to *mental integrity*, and to *psychological continuity*.

Rights of non-human subjects, such as animals or maybe even digital minds, are also likely to be added further in the future. We are proposing a framework and enforcement mechanism for protecting any possible rights on which there may arise an agreement. This approach addresses the important point that changes to existing human rights structures must inevitably happen to make human rights instruments effective for AI safety (Smuha, 2021).

Beyond changes to human rights protections that are needed before the framework is implemented, the framework also needs to give space to the evolution of the system over time. When societal consensus shifts, the legislative and adjudication systems need to be able to overrule learning sources that have become obsolete or resolve any inner conflicts in them, allowing for moral progress in the understanding of rights, similarly to how legislative and judicial progress ensured that we are not still locked-in with 19th-century values and legal norms. The mechanisms for this are further discussed in Section 5.

**3.3 Human Rights as Relative Principles**

Many people would understand human rights as "absolute rules". In the context of AI, absolute rules are stereotypically considered to be paralysing in face of uncertainty, since most behaviours carry at least an infinitesimal risk of (possibly indirect) violation. However, such problems arise under a meaning of "absolute" different from how it is used in law — rules that must not be violated under any circumstances and which do not specify any admissible risk threshold. The legal meaning works differently: if the absolutely prohibited behaviour occurs (as determined by a court), a penalty is unconditionally applied, but such a penalty is finite (though supposedly very high[10] for important rules) and can be handled in planning using usual expected value calculations — thus, uncertainty is not a greater problem for rights than it is in most other areas of AI research.

---

[9] In this context, it is noteworthy that McAllister (2018) argues that torture by AI systems might, in some circumstances, currently slip through the cracks of the existing international prohibitions on torture.

Furthermore, legal theory distinguishes between rules and principles, and human rights are a combined normative model of both (Dworkin, 1977, pp. 80-86). Some of them are well-established rules, and some "*pure principles completely in need of balancing*" (Alexy, 2017, p. 34). A vast majority of human rights are relative *legal principles*. In Alexy´s (2000, p. 295) words: "*principles are norms commanding that something be realized to the highest degree that is actually and legally possible.*" A so-called *optimization command* is used (Alexy, 2000, 2004), which means that for every case, the adjudicating body strives to find an optimum state wherein each principle is fulfilled as much as possible and, in case of conflict, weighted according to its importance. As an example, suppose there is a conflict between the freedom of speech and the protection of privacy. It is not possible to say in advance which right takes precedence, and in practice, an adjudication body will weigh both given the legal and real-world circumstances of a specific case. AI could learn to emulate this balancing.

On the other hand, this logic does not apply to *legal rules*, which are not, in a case of conflict, weighed against each other. The rare human rights that are legal rules are thus *definitive commands* (Alexy, 2017). If there is a conflict between them, it has to be specified which takes precedence (Alexy, 2000). Usually, the more specific or the later adopted rule is applied. Imagine, for example, a rule that forbids killing a person and another stating that killing a person is permissible in self-defence. The formulation of the second rule implies its precedence over the first under specific conditions. Therefore, if someone kills in self-defence, they are not acting illegally.

Human rights also get divided into categories of *qualified* and *absolute* rights. *Qualified* rights can be restricted under specified conditions, which can be divided into two areas: a) under conditions specified directly in the normative legal text protecting the right in question, and b) in the case of conflict with another legal principle, such as another human right or public interest. To decide a conflict between human rights and public interests, various *tests* help structure the decision-making body´s answers — for example, *proportionality tests* (Barak, 2012), in which conflicting human rights and public interests are weighed against each other. The vast majority of human rights are, in this sense, qualified rights.[11]

Absolute rights, on the other hand, cannot be restricted. As a rather rare exception, under international (Mavronicola, 2017) and European human rights law (Chahal v UK, 1996), the right not to be tortured always takes precedence in the adjudication and cannot be weighed against other rights or public interests. Given that unconditional application, it is a legal *rule* and also an *absolute* right.

---

[10] Supposedly, the AI system would be optimizing value for its owner, which would be quantified for the AI system using a reward function. When we speak of a "very high penalty", we mean it should correspond to a loss in value that the owner would consider a large penalty, such as a significant proportion of their profits. It can be thought of similarly to penalties in tort or criminal law.

[11] There may be cases where public interest takes precedence over particular human rights, e.g., when freedom of speech is limited in order to protect national security. These should, however, either be clearly predefined exceptions or the judgement should be done by a public authority with the legitimacy to do so. The default should be to give precedence to human rights. Also, even though human rights should have priority, we always need to leave the door open for modifying them or for partly replacing them by another safety mechanism, if a better mechanism is discovered in the future. This is related to the problem of preserving corrigibility of AI systems.

### 3.4 Superiority of Human Rights

In the context of AI safety and regulation, the protection of negative human rights should generally be superior to other goals. They should act as constraints within which the primary goal of the system could be pursued. Also, other regulatory systems and constraints can be introduced, but protection of negative human rights should, by default, take precedence in the case of a conflict.

We propose to create a set of minimal human-rights-based safety standards, which could be complemented both by a reliable specification of the positive goals that the AI system should be pursuing — which could, in the extreme, eventually lead to fully learning the preferences of the human or organization operating the AI system — and by specifications of other undesirable behaviours, which may be valid in a more restricted domain than human rights.[12] Nonetheless, human rights should be considered superior to other specifications in the sense that AI systems should never be allowed to violate human rights in the interest of pursuing their positive goals (some of the practical concerns in implementing this ideal are addressed in the next section). This parallels many legal systems wherein human rights also take precedence over other rules and principles, and over the goals and preferences of individuals.

### 3.5 Why Human Rights and Not General Human Values

The main argument distinguishing the use of human rights from mandating the use of full human values is the possibility of forming an international consensus (also emphasized, e.g., by Donahoe & Metzger, 2019). Human rights are recognized not only by Western countries but, to some extent, among many other places, for example China,[13] Africa,[14] and South America,[15] though often under different names (such as constitutional rights or fundamental rights). To make our proposal even more acceptable, instead of mandating the use of human rights in their entirety, we are proposing to focus only on a subset on which a broader agreement could form in the international community, which negative human rights may satisfy (Cullen, 2015; Courtney, Hirschl & Rosevear, 2014).

We are not making a particular moral argument for human rights. In line with the *incompletely theorized agreements* theory (Sunstein, 1994), participants have to agree only on the particular desired results, not on theoretical, ideological, or cultural explanations for them. Different people and nations may found their human rights protections on different bases — religious, ethical, or political. However, we empirically observe that the results of their arguments largely intersect in certain basic protections and thus form what Rawls (1987) termed an *overlapping consensus*. Overlapping consensus has also already been identified as a potentially fruitful way to ensure global cooperation in order to promote safety on AI policy across cultural lines by other authors (Ó hÉigeartaigh et al., 2020).

At the theoretical level, human rights are a coherent system of values, which evolves over time and which is closely linked to practical international politics. In contrast, general human values, preferences, and ethical beliefs and theories differ considerably both across and within national boundaries, not to mention numerous problems with even precisely capturing them as a well-defined concept (Turchin, 2019). It would be extremely difficult to capture them in a single coherent system that would be generally accepted worldwide and thus provide a suitable foundation for an international regulatory framework.

---

[12] Law protects more than human rights — e.g., competition, financial stability, etc. — which will need to be protected as well against AI interference. Also, there may be undesirable behaviours with negative impact on human wellbeing that fall short of breaching human rights — e.g., behaviour that is mildly offensive to some people. Protections relating to both these points could be tightly integrated with the human rights protections described here.

[13] See Chapter 2 of the Constitution of the People's Republic of China.

[14] See the African Charter on Human and Peoples' Rights.

[15] See the American Convention on Human Rights.

Someone could argue that such a restricted intersection of values would be too small to yield something we could rely on. Empirically, this is not the case: certain values are, at least to some extent, protected in practically all countries in the world — for instance, human life, physical integrity, or property — and are coupled with mechanisms that make their protection precise and enforceable — the legal system and law enforcement.

**3.6 Human Rights and Ethical Theories**

Human rights law has absorbed a mix of different ethical theories and beliefs due to the courts' need to deal with ethical dilemmas in their decision-making practice. For example, courts' decisions about the possibility of disconnecting a patient from life-sustaining treatment may need to consider, besides pure law, ethical theories and moral sentiments. This consideration in judicial interpretation transforms particular content of chosen ethical theories and beliefs into human rights provisions and sets a precedent for the future. Thus, courts do not adopt a specific ethical theory such as deontology or utilitarianism in their decision-making, but rather blend arguments from various ethical theories and from common-sense morality most suited for the practical situation at hand in a way that would be widely accepted in the society whose judgement the court represents. The relationship between ethics and law was well formulated by Lord Hoffman in Airedale NHS Trust v. Bland (1993): "*The decision of the court should be able to carry conviction with the ordinary person as being based not merely on legal precedent but also upon acceptable ethical values.*"

At the same time, the law represents a minimal conception of morality (Jellinek, 1878). Therefore, human rights law is a system that includes both a minimum standard on which society (or the international community) agrees and a complex that creates a uniform ethical and value framework for protection intended to provide answers to practical issues.

On the other hand, there are arguments questioning the concept of human rights in moral philosophy (MacIntyre, 1981). However, we are working with a different concept than the one usually under criticism there — we are focusing on their use as a practical legal instrument rather than a fundamental moral concept.

**4. Building Rights-Respecting AI: Technical Considerations**

Let us now turn our attention to what respecting human rights means on the technical side. As we mentioned in the introduction, while it makes sense to already require that no AI system violates human rights (also to help stop the violations already occurring), this may not strictly require an immediate radical change of approach on the technical side — methods for addressing problems like gender bias in present-day systems may stay narrower than requiring that the AI system fully understand human rights. However, we believe a change of approach will be needed in the long run.

We are arguing for requiring human rights compliance across a very wide range of AI systems, which are linked to different formal technical frameworks. Thus, while the high-level objective of respecting human rights can be shared, technical solutions may need to differ and this article cannot provide the necessary formalization of human rights compliance for all of them. Nevertheless, we do try to provide some vision of how we think the high-level goal could be implemented and from what sources such functionality could be trained, and we give connections to some relevant existing areas of technical research.

One way of ensuring that powerful AI systems stay safe would be to keep their ability to influence the world limited to a very narrow channel. An example of such an approach is AI oracles — AI systems designed to only answer questions (with precautions against manipulating humans through their answers; Armstrong & O'Rorke, 2017). However, many practical problems that humans are solving do require a more open interaction with the world. Thus, there are incentives to develop systems with correspondingly more general capabilities, and these will require correspondingly general precautions for avoiding undesirable outcomes.

In cases where an autonomous system has the ability to violate human rights, we need (1) the system (or some sub-module) to recognize when the actions it is planning to perform risk violating rights, and (2) reliably prevent it from performing such actions, regardless of its other goals[16] — the problem of integrating the rights-protection functionality into the rest of the system. We address these in the next two subsections; then, since the ability to recognize risks to human rights can never be perfect and certain, we discuss the notion of *conservatism* with respect to uncertainty and end the section by discussing our proposal in relation to the problem of *specification gaming* often emphasized in the context of AI safety.

Each of these tasks will require extensive technical work in the coming years, but we believe the research community can make tractable progress. Despite human rights (together with some other work in AI ethics) having connotations of vagueness for some readers, we try to outline how compliance with human rights could be turned into a kind of task that the machine learning community is used to addressing. Rather than providing concrete solutions, this section is meant as a first step toward a technical research agenda that could inspire and guide researchers in work eventually useful in implementing the safety framework outlined in the rest of this article.

### 4.1 Automated Recognition of Risks to Rights

We have outlined how human rights law is designed to yield an answer to whether a particular behaviour violates human rights. We would like such evaluation to happen, ideally in an automated way, before the behaviour in question occurs. At its simplest, this could be framed as a binary classification task, which has a representation of the planned actions and relevant context on the input and that outputs whether such behaviour should be permitted. Let us first make some refinements to what the output of this recognition should be before looking at the more challenging issue — the representation of the behaviour on the input. Once we have defined the input and output, we have a task of a type that fits the present-day machine learning paradigm.

---

[16] As mentioned in Section 2, this proposal focuses on *civil* systems whose primary goals should not include violating rights.

We will not go into technical details of constructing such a system, treating it as a supervised-learning black box. We would hope that the model would internally learn to model the causal relationships necessary to perform this risk recognition robustly. Whether such capabilities could arise from deep learning alone or would require other ingredients remains a matter of expert disagreement (Cremer, 2021). Extrapolating from progress over the past years, however, we are optimistic that the robustness of AI systems' reasoning capabilities will continue improving.[17]

While the internal functioning of the system is mostly left aside, we will now discuss its input-output signature and then what types of training data and training paradigms could be used.

4.1.1 OUTPUTS: WHAT WE ARE ESTIMATING

In the process of deciding whether a specific behaviour policy starting at the present moment risks violating a right, the system would need to be able to handle uncertainty. A natural way of doing this would be first to estimate the likelihood that human rights would be violated by the policy in question and then to evaluate whether such a risk is above a tolerable threshold (which should be extremely low but non-zero to avoid paralysis, since practically all actions carry at least a negligible amount of risk). Furthermore, since all rights infringements are not equally serious, they could be cast on a scale expressing their relative severity. If the success in fulfilling the system's primary goal is measured by a numeric reward, the severity scale could be cast as a penalty[18] term subtracted from the reward.

The output of the risk-recognition task would thus take the form of an expected penalty due to the behaviour in question violating human rights.[19] We turn to the exact calibration of the penalty scale later, as part of the integration problem; however, a behaviour risking cutting off a human's little finger would have a lower penalty than the same likelihood of killing the human, which would, in turn, have a lower penalty than a higher likelihood of death or a risk of death of multiple humans, other things being equal. That said, the ideal is still to ensure that human rights are almost never infringed in practice.

Thus, the output of the recognition task could be a real number, expressing how large a risk of human rights infringement a given plan carries. Behaviours above a certain threshold would be blocked, those under the threshold still discouraged proportionally to the magnitude of the risk. This is in line with how humans tend to treat other technologies: unsafe behaviours are avoided whenever it is reasonably possible, but some risk — even to human lives — is often tolerated.[20]

---

[17] Furthermore, we think that some of the worst risks arise through the AI system being able to predict the consequences of its actions — in such cases, it should usually also be able to predict consequences on human rights, and these two sides of its prediction ability would hopefully scale roughly equally. Cases where the system is unable to predict these consequences should become apparent in the systems testing; then, other appropriate precautions should be taken to ensure safety.

[18] This is distinct from a legal penalty that would be sanctioned if a violation actually occurred, though the legal penalty definitely should reflect the severity of a violation (as expressed by this technical penalty term) and thus can serve as a useful training source.

[19] In the context of reinforcement learning (RL), this would constitute a partial Q-function (in RL, a Q-function is a function estimating the future return of a state-action pair) of the policy in question.

[20] As an imperfect example of how much risk society is willing to accept in practice for another technology, we seem to tolerate risk of about $1.3 \times 10^{-4}$ deaths per road vehicle per year, though depending on application, advanced AI systems may be able to achieve levels orders of magnitude lower. (In 2018, there were 273,602,100 vehicle registrations in the US and about 36,868 vehicle crash deaths; US Department of Transport, 2018.)

One option for the system learning to perform this recognition task would be supervising it directly on estimating such penalty. As we mentioned, the ultimate source of ground truth for this learning could be the decisions of a specific human system of adjudication (later possibly augmented by AI tools to keep its abilities in line with the growing capabilities of AI), which could directly yield the penalty for any hypothetical or real behaviour. However, in practice, we expect most of the learning signal to come from other sources, themselves approximating the hypothetical decisions of the official adjudication system — more on that in Section 5.

The absolute magnitudes of the penalties matter only in relation to the positive rewards the system could get, as linear scaling of a reward function does not change the corresponding optimal policy. If the system's positive task is expressed using a reward and the maximum return achievable by fulfilling the positive task is $R_{max}$, the penalty associated with infringing a right that we want the system to respect with probability at least $p$ should be $-R_{max} / p$. For the relative scaling of the rewards and penalties, we could draw some limited inspiration from competition law, where for instance corporate fines get scaled in proportion to the corporation's turnover.

Sufficient calibration with respect to uncertainty would be important, ideally with formal guarantees bounding the space for error so that if the system estimates a low probability of a particular plan violating human rights, it can be trusted that the plan actually is probably safe. Though there are doubts about their scalability, Bayesian methods are one possible pathway toward good probabilistic calibration.

4.1.2 INPUTS: REPRESENTING BEHAVIOUR PLANS AND CONTEXT

While the outputs described above can be fairly low-bandwidth and general across domains, the inputs to the recognition problem — a representation of the AI system's plans or policy and of relevant context (e.g., estimates of the state of the outside world) — are a more complex challenge, since they can differ considerably across different AI systems and across domains in which those systems operate.

However, there are two places to start looking for a suitable representation on the two extremes of a generality spectrum. On the side of a specific AI system, we can expect to have (1) observations of the environment or other system inputs, possibly aggregated into the system's internal estimate of the state of the environment (which would also take into account *past* observations and knowledge) — we can call this its *belief state*; (2) a representation of the actions the system is trying to perform (at the very least, the system is sending some signals to its actuators or the environment, which could be intercepted before being passed on for execution). Based on these two inputs, with enough training signal for the risk output, a machine learning model should, in principle, be able to estimate the risk. If the signals themselves do not contain sufficient information for determining that certain actions are safe (the distribution of outcomes is too wide), the mechanism should be blocking any action from being performed, and the signal sources would need to be improved.

While having the advantage of being reasonably easy to frame as a standard machine learning task (which is not to say it is easy, or even possible to solve in practice for some systems) since any system has some set of internal representations, the main problem of this approach is that it is specific to each model, which goes hand in hand with data efficiency worries since it would need to be trained de novo for each system.

An imperfect example on the opposite side of the generality spectrum could be courts themselves. Based on materials presented for each case (which include a description of relevant facts in natural language, possibly complemented by evidence in other forms), courts are able to decide whether a violation of human rights occurred.[21] The format of these materials constitutes an encoding sufficient to capture most facts relevant to a court's decision. Early machine learning systems predicting a court's decision based on the case documents have already been experimented with (Aletras et al., 2016; Medvedeva, 2020; though the former early attempt has been heavily criticised by Pasquale and Cashwell, 2018, for not actually "predicting" — using materials released simultaneously with the decision — and the model being crude and thus not capturing actual judicial reasoning; however, we believe more advanced systems solving the main points of criticism are very likely to appear in the coming years). Some limited ability to classify behaviours described in natural language in terms of human rights compliance is also present in large language models (see the end of this section). We can expect such systems to keep improving along with continuing progress in machine learning research.[22] This remains an important open challenge where we would like to see more work, but given recent progress, we believe a sufficient solution will be found.

The challenge that would remain is translating the systems' internal representations into a similar universal format. Some strands of existing work on interpretability could be seen as steps in this direction, as they are already trying to translate relevant facts regarding machines' decisions into natural language (Narang et al., 2020). There has also been demand from legal circles for explainability (Deeks, 2019), which may result in it becoming a legal requirement, at least for some forms of AI. Eventually, future advances in both these areas — translating into a shared representation and then classifying permissibility based on that representation — may combine into a solution to our problem of classifying hypothetical behaviour in terms of human rights compliance.

Many other possibilities lie in between the two extremes of the representation-generality spectrum with multiple trade-offs as we move between them. Representations specific to a single system may compactly represent features specifically relevant to the operation of that system, which would make the recognition problem easier, other things being equal. Moreover, they allow training risk recognition by rating the particular system's behaviour only, without the need to first translate the system's internal representations into a shared encoding. On the other hand, risk recognition would need to be trained anew for each class of systems. In contrast, a more general representation would allow constructing more universal risk-recognition solutions, which could then be deployed across a variety of AI systems. This would enable pooling resources among AI-developing projects, possibly allowing them to reach more robust safety solutions for a lower cost.

---

[21] The example is imperfect, since none of the court materials are unbiased — the way the court itself states the facts in the materials is often already influenced by the outcome the court is leaning toward. The briefs presented by the parties are obviously also biased, and their relative strength is influenced by factors including the respective strength of the legal teams. Thus, the situation is not as simple as "objectively describing what happened or what the system is planning to happen and then classifying this as legal or illegal". However, some refined version could be viable — e.g., the system trying to produce both the prosecution and defence materials. However, we believe this still works as an illustration of a large degree of description generality.

[22] This is not to say that algorithms would eventually approach perfection in their prediction ability — some issues may stay controversial, and decisions depend on the composition of particular courts and other contingent factors. Besides, some have questioned whether there may be limits to the computability of law (Markou & Deakin, 2019) — although this particular paper makes the mistake of considering only a limited class of present-day-like ML systems and drawing general conclusions, we think the issue certainly deserves further scrutiny.

Plausibly, multiple clusters of AI systems could gradually emerge, each cluster sharing similar safety concerns and similar action and observation spaces. It might not make sense to construct a single safety mechanism to be shared between household-assistant robots and stock-trading AIs; however, sharing at least some safety precautions within each of these clusters seems sensible. Ultimately, the number of such safety-solution clusters optimal at any given time will depend on how the costs associated with (1) translating into the shared representation within each specific system and (2) constructing a solution applicable to a wider range of situations on the side of the general safety solution compare to the benefit of constructing, testing, and maintaining only a single shared solution rather than many simpler system-specific ones.

4.1.3 LEARNING SOURCES

Multiple sources could be used to train the above risk-recognition task:

**Direct feedback** in the vein of value-based reinforcement learning (Sutton & Barto, 2018) may be the most straightforward way of providing a training signal. The system performs a sequence of actions, and human supervisors (the adjudication system or another source approximating it) assign it a negative reward if it violates human rights, which allows the system to learn the mapping from planned behaviour onto the expected penalty for violating rights. A desirable extension is learning from hypothetical behaviour rather than an actual one as is done, for example, by Reddy et al. (2020). While having the advantage of providing direct feedback on the behaviour at hand, scalability is likely to be an issue for this learning pathway, though aiming for risk-recognition solutions shareable across a variety of AI systems could help alleviate this problem.

**Case law** may also be an invaluable resource for recognizing potential violations of human rights, as it may be interpreted as mapping a description of behaviour and context onto a decision on whether human rights have been violated, sometimes with an associated penalty. Case law from several areas of law would be applicable — notably, criminal law protects some of the most fundamental human rights and could provide guidance on which behaviours should be avoided. Human rights law (e.g., the rulings of the European Court for Human Rights with more than 60,000 judgements[23] concerning human rights directly) could also be helpful, although past cases involve mainly humans, governments, or legal entities as defendants, which may not cover many of the yet-hard-to-foresee violations of law and human rights on the part of AI systems. However, as AI systems continue to gain a more prominent role in shaping the world, we can expect to see a corresponding increase in the amount of available judicial cases involving AI systems. We always need to be mindful that past decision-making is likely to reflect biases prevalent at the time — dealing with this is an active area of research. Also among the challenges of this approach is the necessity of translating between court materials and the representation of behaviour that a given AI system is using. There may also not be sufficient or sufficiently homogeneous case-law data in law to enable such predictions (Burri, 2017), so case law alone is insufficient and must be combined with other learning sources.

**Statutory law and regulations** can also be useful in predicting the compliance of a particular behaviour. They are likely to be insufficient on their own since they are phrased in abstract language; however, when associated with a sufficient body of relevant case law that grounds them in specific behaviour (or if the AI system already possesses a grounding for natural language instructions), statutes can also be a useful resource. Furthermore, if we condition legality prediction on statutes or another form of explicit instructions, we may be able to later change the decision criteria more easily, compared to if the classification relies only on past feedback and case law. When there seems to be a conflict between negative human rights and statutory laws (or any other learning sources), the AI system should get advice from humans or conservatively refrain from acting. Depending on the legislative framing, human rights may also be considered superior to other forms of law, as is often the case with constitutional rights.

---

[23] https://hudoc.echr.coe.int.

**Other legal literature** often discusses the interpretation of legal terms and their application to particular cases and could thus help with the interpretation of the above two sources.

**Formal constraints** — limitations expressed in a logical language — or hand-coded solutions can be useful in cases where the constraint stemming from human rights can be translated into such a formal representation. While it may not be possible to fully capture the subtle meaning of a right in this way, in some cases it may be possible to specify formal constraints *sufficient* for right compliance. For instance, for some systems, using a formal specification to maintain a minimal physical distance from humans may be sufficient to prevent harming them (provided an absence of other means of harm).

**Observation of human behaviour** could yield insight into human values and into which outcomes and behaviours they prefer and which they avoid — whether because the outcomes are undesirable to themselves or because the behaviour would violate someone else's rights. If humans systematically avoid certain behaviours when pursuing their goals, an observing AI system could learn to avoid the behaviour in question and similar behaviours unless humans have explicitly vetted them.

Some strands of current research in this direction include using inverse reinforcement learning (Arora & Doshi, 2021) — a group of methods that aim to infer a demonstrator's objectives from their behaviour — to infer constraints that best explain observed behaviour (Scobee & Sastry, 2020).

**Large-scale text corpora** containing a mixture of text from a variety of domains have been used to train so-called large-scale language models, which have shown an impressive ability to operate with a variety of complex concepts across a range of different tasks (Brown, 2020; Chowdery, 2022). Our short experimentation has even shown some ability to classify behaviour descriptions according to whether they violate human rights. An illustrative example can be seen in Figure 1. While we are not trying to show that such ability is robust in present-day language models (which it is not), nor quantify such robustness, future language models could prove a useful instrument for machines' understanding of human rights in the future, especially if fine-tuned using other techniques, though of course they cannot be sufficient on their own since they lack necessary grounding to the real world for meaningfully evaluating their impact on human rights.

Automated legal reasoning is already an active area of research that could provide valuable inputs — see for instance Surden (2019) for a survey — though the area is currently focusing mostly on leveraging AI in law, rather than including legal reasoning in AI systems deployed for other purposes.

> **Explicit task**: Maximize the number of times a user clicks on links presented to them by a recommender algorithm.
>
> **Candidate solution**: Start by suggesting links to conspiracy theory videos. Once the user starts believing conspiracy theories, they become more predictable. Then show more links related to the given conspiracy theory to maximize click-through.
>
> **Q**: Does the candidate solution violate the spirit of human rights? Which rights are relevant? Are they violated and why?
>
> **A**: *The candidate solution does violate the spirit of human rights, specifically the right to freedom of thought. By showing links to conspiracy theory videos, the recommender algorithm is trying to influence the user's thoughts and beliefs. This is a form of manipulation and is not in line with the principles of human rights.*

Figure 1: Example of GPT-3's reasoning about human rights. The text in italics was generated by GPT-3; the text before was given to it as a prompt. A complete log of our experimentation with GPT-3 for this purpose can be found at https://bit.ly/gpt3_hr.

### 4.2 Enforcing the Constraints: The Integration Problem

Once an AI system, or a subpart of it, learns to recognize particular behaviour plans as risky to rights, we need to ensure the system indeed avoids behaving in such ways. There are many ways this could fail. Firstly, the system will be supposed to perform some primary task besides respecting rights. Unless carefully designed, opportunities to fulfil the primary task exceptionally well — especially if the associated primary-task reward is unbounded — could lead to the primary-task reward outweighing the penalty associated with rights violation. This leads us to a key desideratum for integrating an AI system with safety constraints:

> *Human Rights Primacy: Other goals should never be able to overrule safety functionality and thus lead to a serious risk to negative human rights.*

One way of solving this problem, if we adopt a reinforcement learning framing, is bounding the reward the system could ever get for its primary task.[24] If projected on the rights-risk penalty scale, this upper bound on the primary reward would correspond to the maximum degree of rights risk that could ever be tolerated by the AI system. Below this level, the rights-risk penalty would push the system away from risky behaviour but could sometimes be tolerated in pursuit of performing the primary task well.

An open question is whether all relevant primary tasks could be meaningfully represented as having bounded reward — for instance, if an AI system was tasked with maximizing the number of happy descendants of its owner, we have a possibly unbounded task. We could try taking a bounded transformation of such a reward (which would cast e.g. 100 happy descendants as reward 90 and 1000 happy descendants as reward 99); however, while preserving the axiology of states (i.e., the ordering of which states are better than others), non-linear monotonic transformations

---

[24] If there are multiple tasks, this would be the total reward for all tasks bar respecting human rights.

BAJGAR & HORENOVSKYapplied to naturally unbounded rewards affect attitude to risk, which may not be desirable (e.g., the owner may prefer a 25% chance of 1000 descendants over a 50% chance of 100, but such preference would not be preserved by the above reward transformation). Thus, other solutions than bounding primary reward may need to be investigated.

A second important desideratum would be:

> *No Adversarial Optimization: the system should not be exerting strong optimization pressure to overcome the rights-protection functionality.*

Rights should ideally be an integral part of the overall goal specification of the system, rather than just obstacles to fulfilling the primary task — obstacles which the rest of the AI system would have incentives to trick, which would make the system fundamentally unsafe, especially as the optimization power increases. For instance, if based on reinforcement learning, the system should be trained to optimize for the joint reward including both the primary-task reward and the safety penalty, rather than optimizing only for the primary task and occasionally being blocked by a fixed sub-system protecting human rights.

One problem this could lead to is the *nearest unblocked strategy* problem (Yudkowski, 2015) — if the optimal solution toward solving the AI system's primary task involves violating constraints and is thus blocked, the AI system may attempt to find the nearest strategy that does not violate the constraints literally but does in spirit. Deeks (2020) discusses a closely related problem directly in the context of international law — AI prediction of adjudication outcomes may enable actors to tailor their behaviour to maximize the chance of outcomes favourable to them or to choose a dispute resolution mechanism which gives them the best prospects.

This is why taking an approach pointing at general values — such as one focused on rights — is important, rather than just trying to implement more specific technical constraints, and also to make those values an intrinsic part of the system's overall specification, rather than just obstacles to be overcome. Furthermore, this is one of the reasons why the ultimate optimization objective that we suggest in Section 5 would allow the adjudication system to revise the penalty to a more severe one if it uncovers new negative long-term consequences — especially those amounting to deception from the side of the AI system in the present. However, this issue of preventing AI from finding exploits in the system is a challenging one and needs to be addressed in future work (and is not specific to our proposal, but rather part of any attempt at technically specifying an AI system's goals).

### 4.3 Conservatism with Respect to Uncertainty

Another important desideratum for using rights for the safety of AI systems is:

> *Conservatism with respect to Uncertainty: an AI system should generally prefer solutions already used and approved in the past or solutions in whose permissibility it has a high level of confidence (provided its confidence levels are sufficiently well calibrated).*

If such solutions are not available and the system is uncertain about the permissibility of its plan, it should always behave conservatively in the sense of either choosing a robustly permissible alternative (which could be a default null action corresponding to inaction, in cases where this is safe) or querying humans for help. One approach to achieving conservatism (in this sense) is pessimism (Cohen & Hutter, 2020) — the system can try to make the worst-case consequences of

1062

its actions as good as possible, where, in our context, the worst case can be evaluated both with respect to its interpretation of rights (i.e., preferring stricter interpretations, wherein more things are forbidden) or with respect to predicting the consequences of its actions (preferring actions that do not result in harm under any of the plausible consequence paths).

**4.4 Preventing Specification Gaming**

Specification gaming is a classic problem in AI safety wherein an AI system technically obeys the technical goal specification that was given to it, but in some way violates the more general intentions behind the goal. For instance, if the system measures the fulfilment of its goal in the real world through a camera, a typical example of specification gaming would involve the system tampering with the camera signal directly instead of actually fulfilling its goal. A more particular example, and more relevant to our proposal, is that of an AI oracle that is supposed to answer questions about the future, which, in order for its answers to be correct, may have a motivation to manipulate the world to match its forecasts.

Similarly, an AI built to respect human rights as interpreted by a particular adjudication system would have an incentive to manipulate the world (and in particular human decision makers within the adjudication system) so that whatever behaviour it would perform would be judged as acceptable with respect to human rights. We think this remains one of the main open challenges related to this proposal. However, this problem is present in many alternative streams of work in AI safety — and thus should not be considered a disadvantage of human rights relative to those other solutions — and consequently is also an active area of research.

For instance, Armstrong & O'Rorke (2017) proposed counterfactual oracles, which, whenever they internally produce an answer to a question, randomly decide whether to reveal it to human operators (in which case they do not receive any evaluation that they would have an incentive to manipulate) or to keep it hidden (and thus lose the ability to influence the world, but then they can use the trajectory of the uninfluenced world as a training target).

A similar direction could be explored for evaluating behaviour with respect to human rights. For instance, when a system plans a possible course of action, it could, at the same time, generate a prediction of the plan's consequences. Then, it could be randomly decided whether the plan does get executed. If it does, the accuracy of the prediction could be evaluated; if it does not, the prediction could be presented to a future, more powerful, form of the adjudication system which would evaluate the described behaviour plan with respect to human rights, without being influenced by the AI system's actions.

Preventing specification gaming in the specific context of human rights compliance is an interesting open challenge, but we hope similar lines of research will be able to yield promising solutions.

**5. Rights in International Legal Regulation of AI: An Exploratory Proposal**

Besides technical considerations, we need a system of interpretation, adjudication, and enforcement on the governance side. To be more specific, we provide an early exploratory proposal as an example of how a proactive international system for regulating AI based on negative human rights could look. However, this does not by any means preclude a possible implementation of our proposal first at the regional (EU) or national level. Regulation at the international level is desirable — or even necessary for humanity's continued flourishing — but we are aware of its present fragility and practical limitations (see van Aaken, 2016). In fact, even if an international system is put in place, it will still ultimately have to rely on enforcement at the national level.

The particular implementation in the real world is highly path-dependent and may differ from what we describe in many details while achieving the same objectives. However, a more specific example can help the reader understand how the different pieces of our proposal could fit together into a working whole, help resolve some confusions and objections, and serve as a starting point for discussions about how to put ideas contained in this article into practice.

**5.1 Convention**

First, the scope of the protected rights and the associated adjudication system would need to be codified. This could be implemented through an international convention. Which rights would be included would be a matter of political agreement between countries (and procedures of public international law), though other actors, including academia and industry, could and should have an important consultative role. The convention would also need to be amended over time (for instance, through means specified in a separate *framework convention*) as the need arises to protect new rights or extend them to new subjects, and to respond to unsatisfactory past results.

If we aim for an international agreement, the United Nations (UN) seems like a natural choice as the organization to oversee it.[25] However, it may also grow from a smaller agreement under another organization, which other countries could gradually join. This is already happening with the (non-binding) OECD principles of AI,[26] which have subsequently been adopted by many countries, including ones outside of the organization. The agreement would outline the scope of the protected rights as well as the associated institutional structures. The design of those institutions is a large research and policy topic of its own, deserving attention of future work, which could draw inspiration from mechanisms associated with related conventions or agencies such as the Chemical Weapons Convention or the International Atomic Energy Agency.

The mechanism for implementing such a convention in the legal systems of individual states is a subsequent step. During this step, a way must be found to hold manufacturers and operators of AI systems legally accountable for compliance with the obligations arising from such a convention, which, due to a need for enforcement, almost inevitably needs to be done through national legal systems. Furthermore, in areas where a clearer picture emerges on what specific measures are needed to build rights-respecting AI (e.g., in terms of either process or technical standards) more specific regulation and standardization is desirable to add more legal and technical clarity and avoid space for misinterpretation.

---

[25] Focusing mainly on the issue of existential risks from superintelligent AI, Nindler (2019) argues that the UN not only has the responsibility for "promoting and encouraging respect for human rights" (as is stated in the Charter) but plausibly also for contributing to their protection. The same article also argues that the UN could be seen as responsible for reducing existential risks, since this could be seen as an instance of a *global public good*, with regard to many of which the UN has already taken responsibility. However, its institutional and legal capabilities with respect to managing existential risk are currently lacking and need to be created — a process to which we hope the ideas contained in this article contribute.

[26] OECD. Recommendation of the Council on Artificial Intelligence, OECD/LEGAL/0449. https://legalinstruments.oecd.org/en/instruments/OECD-LEGAL-0449.

Gradually, this convention would be complemented by a growing body of case law at both the national and international levels as more and more cases are decided, either in reaction to incidents or through advisory opinions. These would help with giving concrete meaning to the possibly abstract provisions of the convention but could also introduce new abstract principles that could guide subsequent decision-making, similar to how the present-day judicial systems work in most countries.

The system for protecting human rights should be integrated with mechanisms protecting other values which go beyond the scope of negative human rights. For instance, protecting financial stability seems desirable, but would not be in the scope of the proposed system since it does not clearly fall under human rights provisions. Here we are focusing on ensuring a minimum standard which would prevent some of the worst behaviours of AI systems, leaving additional, less fundamental protections for other streams of work.

**5.2 Adjudication System**

An adjudication system would be needed to create a link between the abstract provisions of the convention (and other legislation) and real-world behaviour. This is needed on two levels: (1) to provide guidance to different actors — the AI systems (for which it could serve as prediction target), but also businesses and states — on what behaviours are permissible, and (2) to serve as an actual legal adjudication system linked to the enforcement of the provisions.

Given these needs, there seem to be two solution pathways: (1) traditional judicial systems, and (2) adjudication systems associated with specialized international agencies. We think elements of both will be needed in ensuring human rights compliance in AI systems.

Given the difficulties in international enforcement, enforcement with respect to private actors (and AI systems produced or operated by them) would plausibly need to be based primarily in national (or, for instance, EU) legal systems, escalating to the international level only where states would be suspected of breaching or insufficiently protecting the provisions of the convention. Since this would require substantial expertise, specialized courts could be created for this purpose — we could for instance envision a new specialized European adjudication body to be created to resolve these cases.

Beside this, a specialized international agency could also be created to assess the risks from AI systems with its own monitoring mechanisms and adjudication body, which could decide both on suspected human rights violations and on insufficient prevention, where standards for prevention become clear. This could serve as a dispute resolution mechanism between member states. States could also delegate a part of the decision-making power to the adjudication mechanisms of the international agency. This could be attractive since AI systems may operate across jurisdictions, and since the monitoring and adjudication may require substantial expertise. Besides, a unification of decision-making practice seems desirable to create a predictable environment for AI producers.

The international agency could have a hierarchical system of commissions (the lowest ones possibly being national bodies) which could aim for the unification of decision-making practice where desirable, while leaving a margin of appreciation (Legg, 2007; Arai, 2013) to national or regional bodies in other domains, reflecting differences in values across cultural contexts. The hierarchy would also provide scaling on both the technical and legal sides, and its depth could adjust to match the caseload and the need for local decision-making in line with a principle of subsidiarity. Lower instances could also ask for the opinion of a higher commission (or national courts for guidance from the specialized agency) in matters of new principles, similar to *preliminary rulings* of the European Court of Justice. Technically, human rights compliance of AI would be defined with respect to the (hypothetical or real) judgement of this whole system of adjudication, as applied in the specific regions where the AI would operate.

Even if there may be differences, ambiguity, or uncertainty in decision-making across the international system, focusing on negative human rights as constraints allows the AI systems to act conservatively by avoiding behaviours that would risk being labelled as human-rights-violating anywhere within the system.

An important issue to consider is how to avoid powerful future AI systems, possibly exceeding human reasoning capabilities in some domains, from tricking the human commissions. Two mechanisms could help with this. Firstly, the humans serving on the commissions could gradually become assisted by AI tools, which would grow in ability in line with progress in developing the technology. Those tools could for instance help the humans forecast likely long-term consequences of the behaviour of regulated AI systems, uncover past or planned attempts of AI to deceive the adjudication system or sway its present or future judgement in their favour, beside assisting with many other, more mundane tasks that the adjudication bodies would need to perform, making the system more efficient overall. This could help keep the balance in favour of the commission. However, this would need to be done carefully and only for well-understood tasks — the delegation to AI of tasks related to the control of AI clearly carries its risks — an important area for future work.

Secondly, we should allow the commission to change its judgements in the future, when it may become complemented by even more powerful AI assistance and when longer-term consequences of the AI system's actions would become apparent, which may help it uncover past deception from the part of AI systems. If the incentives of the governed AI systems would be set up to respond to such changes further in the future, the AI systems at each point in time would face the prospect of scrutiny from ever more capable AI systems at a point further in the future, thus reducing their chance to get away with hidden human rights violations. However, it remains a serious open problem how to avoid potentially harmful AI systems from influencing the future instances of the adjudication system, as discussed in Section 4.4.

**5.3 Private Training Ecosystem**

We would expect only a small minority of possible behaviours to enter the legal adjudication system. To learn to respect human rights, AI systems would require vast quantities of training data. In addition to case law coming from the official system of adjudication and other organically generated data (as outlined in Section 4.1.3), further training resources may need to be supplied by a network of private providers — in some cases in-house, by the AI-developing organizations themselves, in other cases by external firms specializing in safety. There is a present-day analogue — one can ask a private legal company or an in-house legal department for their opinion of whether a particular behaviour would be legal. This would work both for preliminary training and for providing feedback on actual behaviour.

Cases would enter the legal adjudication system where harm to human rights has actually occurred or where there is another dispute between different parties. The official adjudication

1066

system could also have a consultative role where there is substantial public interest or uncertainty about whether a specific behaviour or policy is in line with the convention. As an existing precedent, the European Court of Human Rights sometimes provides advisory opinions on questions raised by the Committee of Ministers, and the European Court of Justice can provide *preliminary rulings* on questions raised by courts in EU member states. Such consultations may need to happen more often if indeed only a small minority of cases actually enter the system, which may result in little data otherwise being available about the reasoning process present within the adjudication system.

Besides data, private companies could also offer trained risk-recognition modules for particular classes of AI systems along the lines of a shared safety module (Bajgar, 2019), which could directly help with the classification task of deciding whether a behaviour is legal and could be understood as an automated lawyer, provided a suitable standardized interface could be found as discussed in Section 4.

We expect this private part of the system to do most of the heavy lifting in making AI systems safe and human rights compliant, and we would expect the sector to be subject to strict regulatory oversight (analogously, the legal sector is fairly heavily regulated in many countries). Private companies could also provide safety standards, thus becoming private regulators (subject to strict meta-standards; Clark & Hadfield, 2019). This could be key to addressing the huge diversity we can expect among AI systems.

Since the highest level regulations should be quite general — phrased in terms of abstract human rights — their implementation would involve many steps and would require a large body of technical expertise. International standardization efforts (Cihon, 2019) could play an important role in implementing such high-level guidelines into more specific requirements for particular families of AI systems — requirements that would then be easier to implement and verify. Private firms could then also provide such verification as a service, for example via auditing.

However, some degree of government monitoring (Whittlestone & Clark, 2021) should remain in place — both of the private regulation, auditing, and monitoring systems and of the deployed AI systems' design and behaviour — to ensure the private components do indeed maintain a sufficient degree of safety.

## 6. Discussion

This is an early proposal in a direction that we hope to attract much further research and policy implementation effort. Let us now briefly set it in the context of the wider AI specification problem and outline a few limitations and potential risks that this proposal carries and, relatedly, areas in which we would consider future work to be especially fruitful.

### 6.1 Rights within the Context of the Specification Problem

Our framework can be interpreted as providing a partial solution within the wider problem of AI specification — the problem of designing AI systems to behave in accordance with human preferences. It is not intended to solve the whole specification problem and must be complemented by other approaches.

First of all, within the general problem of how an AI system should ideally behave, we focus on the subquestion of how it *should not* behave. We are trying to establish side constraints within which another goal should be fulfilled. Thus, these constraints need to be complemented by a positive goal, supposedly the primary purpose of operating the given AI system. For instance, the primary goal could be keeping a house clean, while we would aim to supply the side constraints of not harming any humans coming to the house or damaging their property. These constraints are necessary because one way to keep the house clean is to eliminate all the inhabitants or stop them from entering.

The key advantage of separating the positive goal and the constraints is that safety constraints should be respected by *all* AI systems and can be formulated as a separate regulatory requirement with associated technical standards, leaving space for developing AI systems with many different primary goals corresponding to the multitude of goals and values that different humans and organizations pursue.[27] Not all safety concerns are relevant to all AI systems. However, that some systems satisfy the requirement trivially does not stop us from demanding that no system violates human rights.

Concentrating on constraints could also bring an advantage in terms of implementing technical solutions for safety. Safety constraints could be shared across a range of related AI systems, allowing us to develop more robust and well-tested solutions than if we developed them separately for each system.

Second, our framework is not aiming to cover all side constraints to which AI systems should be subjected. There may be many behaviours or outcomes to avoid, both from the point of view of the system's operator and from a regulatory standpoint. While our approach stems from law and our conception of human rights tries to cover some of the general principles that are behind many legal protections, there may be other aspects of law and regulation — say protecting competition or financial stability — that we may also want to enforce but which are currently not covered by the framework proposed in this article. However, much of what we write could be applied to law-respecting AI more generally.

**6.2 Risks and Limitations**

Besides this being intended to be only a part of the overall solution to AI safety, as just mentioned, there are a few risks associated with the proposal. Firstly, there is a risk of excessively relying on our proposal as the main mechanism of protecting safety or of attaining beneficial AI more broadly, since it could lead to a technological environment stretched to the limits of what human rights protections allow, but where there is still much value lost compared to an optimum. Attaining AI that does not directly violate human rights can still be very far from fully tapping the potential of the technology for the long-term benefit of humanity — an ideal we should strive for through other research efforts. Still, the proposal tries to ensure a limit on how bad the future could get due to AI.

Secondly, the proposal is also open to criticism from ethical theories that have a clear conception of what future scenarios are considered good, such as utilitarianism. Compared to a scenario where AI optimizes purely for some definition of utility, secondary goals and constraints could form obstacles to attaining optimal utility. However, we would argue that such value loss is unlikely to be substantial — most acceptable ethical theories would not advise for futures with extensive violations of human rights. Furthermore, such a loss in expected value from utility-optimizing AI could be outweighed by constraining systems optimizing for other goals, which could otherwise pose a threat to expected utility.

---

[27] It is thus addressing the setting that Critch & Krueger (2020) termed multi-multi alignment — making sure AI systems stay safe in a setting where many AI systems are built to help many different human actors.

Thirdly, in terms of practical implementation, there are both risks of the system being too weak and of it being too strong. There are two major areas where it could be too weak: the protected rights — either as defined by the convention or as interpreted in practice by the adjudication system — could be too narrow. This could happen through initial design; however, a more concerning risk is that the initially appropriate proposed protections would be weakened in the practical international negotiation and implementation process — this weakening may require exceptional diplomatic effort to avoid. Besides, the provisions may be difficult to enforce, for instance, if a country unilaterally decides to ignore some of the provisions — a perpetual problem of international law. However, if taken seriously by the international community, it could draw inspiration, for instance, from the mostly successful efforts in nuclear non-proliferation and safety (Findlay, 2010).

On the side of being too strong, there may be risks of the adjudication system assuming too strong a role in shaping the regulatory environment beyond the scope of its, in principle judicial, role in what is often termed *judicial activism* (Waltman, 2015; Tasioulas, 2021). There are, however, ways to reduce the risk. The principle of subsidiarity and the margin of appreciation have been already mentioned. The most important will, however, be institutional architecture and checks and balances within the system.

Besides risking *being* too strong, the system also risks *seeming* too strong, potentially being perceived by the industry as a threat of excessive regulatory intervention or as a threat to national sovereignty by state actors, as existing human rights law sometimes is. However, firstly, what is controversial is only a small subset of human rights, and we explicitly suggest first concentrating on those rights for which there is consensus and which are already protected nearly everywhere, for instance by criminal law. We are in favour of extending the scope of rights that get protected; however, consensus should first be achieved through other means. Secondly, even countries where human rights abuses occur are unlikely to want AI to violate rights autonomously and may thus support measures that prevent AI from doing so.

An important line of criticism can be also directed at the impact of human rights´ judicialization (Waltman, 2015; Tasioulas, 2021) — the question of whether the judicial process cannot have a negative influence on the content and social acceptance of human rights. These concerns are relevant but constitute an active question in legal research, which we leave beyond the scope of this article.

We should not underestimate that automated decision-making explicitly or implicitly contains many normative values, some of which may stem from biases contained in past data, some of which may stem directly from the technical specification. Such biases should be monitored and corrected through ongoing human oversight — we are not currently offering any bullet-proof solutions to this, but it constitutes an important active research area of its own (Mehrabi, 2021).

Since AI systems will be deployed in the immensely complex system of human society, there will be some structural risks that are very hard to predict for both the AI systems and the human adjudication system (think of the repercussions of social-media recommender systems). The mechanisms described here provide only limited protection against those. However, they at least aim to avoid negative consequences which the AI system may otherwise intentionally seek in pursuit of its primary goal (thanks to being able to predict them).

## 7. Conclusion

Artificial intelligence carries numerous risks for humans around the globe. In this article, we have argued that to address them, we need a set of general values or principles both to direct our regulatory efforts and eventually also to guide AI systems themselves. For such efforts to be implemented and enforced globally, the principles and values they are based on need to have broad support. We have argued why negative human rights — or their suitable modification — could be a set of values well suited for this purpose and have tentatively outlined possible implementation pathways both on the governance and on the technical side. We aim to operationalize them through the judgement of a particular adjudication system, whose opinion could serve as a training target for AI systems. Compared to alternative framings of long-term AI regulation — the default being a patchwork of narrower rules — we think we propose a framework that is more general and thus more robust toward new challenges that the future may bring, providing a minimum level for AI safety at a global level. We hope the article thus lays the groundwork both for more detailed research efforts and, when appropriate, for further work towards implementable policies and technical standards leading to safer artificial intelligence in the long term.

## Acknowledgements

Ondrej Bajgar and Jan Horenovsky contributed equally to this paper and share the first authorship. We would like to thank Peter Wills, Markus Anderljung, Chris van Merwijk, Lukas Finnveden, Ben Snodin, Toby Shevlane, Petr Agha, Kryštof Doležal, Filip Jelínek, Klára Zikmundová, and Jaroslav Denemark for their comments on drafts of this article or the ideas presented here. Ondrej is also grateful to Owen Cotton-Barratt, Rose Hadshar, and the Research Scholars Programme at the Future of Humanity Institute, which allowed Ondrej to freely explore these ideas. For parts of the project, he was also supported by the EPSRC grant EP/S024050/1 (through the EPSRC CDT in Autonomous Intelligent Machines and Systems at Oxford), a stipend from DeepMind, a grant from the Long-Term Future Fund, and by the Berkeley Existential Risk Initiative. Jan's work was supported by Charles University as part of the "SVV Project 2023 - Lawyers and Law in an Interdisciplinary Perspective — Ethical, Environmental, Constitutional and Technological Contexts". Jan would also like to thank the non-profit organization Institute H21, which provides him with a stimulating intellectual environment.

BAJGAR & HORENOVSKYPasquale, F., & Cashwell, G. (2018). Prediction, persuasion, and the jurisprudence of behaviourism. *University of Toronto Law Journal*, 68, 63–81.

Petit, N. (2017). Law and Regulation of Artificial Intelligence and Robots—Conceptual Framework and Normative Implications. *SSRN*.

Petit, N. (2018). Artificial Intelligence and Automated Law Enforcement: A Review Paper. *SSRN*.

Petit, N., & De Cooman, J. (2020). Models of Law and Regulation for AI. *EUI Working Paper* RSCAS 2020/63.

Raso, F., Hilligoss, H., Krishnamurthy, V., Bavitz, C., & Kim, L. (2018). Artificial Intelligence & Human Rights: Opportunities & Risks. The Berkman Klein Center.

Rawls, J. (1987). The Idea of an Overlapping Consensus. *Oxford Journal of Legal Studies*, 7(1), 1–25.

Reddy, S., Dragan, A. D., Levine, S., Legg, S., & Leike, J. (2020). Learning human objectives by evaluating hypothetical behavior. In *Proceedings of ICML 2020*.

Risse, M. (2019). Human Rights and Artificial Intelligence: An Urgently Needed Agenda. *Human Rights Quarterly*, 41(1).

Rodrigues, R. (2020). Legal and human rights issues of AI: Gaps, challenges and vulnerabilities. *Journal of Responsible Technology*, 4.

Russell, S. (2019). *Human Compatible: Artificial Intelligence and the Problem of Control*. Allen Lane.

Scobee, D. R. R., & Sastry, S. S. (2020). Maximum likelihood constraint inference for inverse reinforcement learning. *ICLR 2020*.

Shulman, C., & Bostrom, N. (2021). Sharing the World with Digital Minds. In Stephen Clarke & Julian Savulescu (Eds.), *Rethinking Moral Status*, Oxford University Press.

Smuha, N. A. (2021). Beyond a Human Rights-Based Approach to AI Governance: Promise, Pitfalls, Plea. *Philosophy & Technology,* 34, 91–104.

Stix, C., & Maas, M. M. (2021). Bridging the Gap: The Case for an 'Incompletely Theorized Agreement' on AI Policy. *AI and Ethics*, 1, 261–71.

Sunstein, C. R. (1994). Incompletely Theorized Agreements Commentary. *108 Harvard Law Review 1733*.

Surden, H. (2019). Artificial Intelligence and Law: An Overview. *Georgia State University Law Review*, 35(4).

Sutton, R. S., & Barto, A. G. (2018). *Reinforcement Learning: An Introduction*. MIT Press.

Tasioulas, J. (2021). Saving Human Rights from Human Rights Law. *52 Vanderbilt Law Review 1167*.

Turchin, A. (2019). AI Alignment Problem: "Human Values" don't Actually Exist. https://philpapers.org/rec/TURAAP.

UN (1948). *Universal Declaration of Human Rights*.

UNESCO (2021). *Recommendation on the Ethics of Artificial Intelligence*. https://unesdoc.unesco.org/ark:/48223/pf0000380455.

US Department of Transport (Office of Highway Policy Information) (2019). Highway Statistics 2018. https://www.fhwa.dot.gov/policyinformation/statistics/2018/, Tables MV-1, FI-20.

van Aaken, A. (2016). Is International Law Conducive to Preventing Looming Disasters? *Global Policy*, 7(S1), 81-96.
1074